# XACML Extension for Graphs: Flexible Authorization Policy Specification and Datastore-independent Enforcement

Aya Mohamed[1,2][a], Dagmar Auer[1,2][b], Daniel Hofer[1,2][c] and Josef Küng[1,2][d]
[1]*Institute of Application-oriented Knowledge Processing, Johannes Kepler University Linz, Linz, Austria*
[2]*LIT Secure and Correct Systems Lab, Johannes Kepler University Linz, Linz, Austria*
{aya.mohamed, dagmar.auer, daniel.hofer, josef.kueng}@jku.at

Abstract: The increasing use of graph-structured data for business- and privacy-critical applications requires sophisticated, flexible and fine-grained authorization and access control. Currently, role-based access control is supported in graph databases, where access to objects is restricted via roles. This does not take special properties of graphs into account such as vertices and edges along the path between a given subject and resource. In previous iterations of our research, we started to design an authorization policy language and access control model, which considers the specification of graph paths and enforces them in the multi-model database ArangoDB. Since this approach is promising to consider graph characteristics in data protection, we improve the language in this work to provide flexible path definitions and specifying edges as protected resources. Furthermore, we introduce a method for a datastore-independent policy enforcement. Besides discussing the latest work in our XACML4G model, which is an extension to the Extensible Access Control Markup Language (XACML), we demonstrate our prototypical implementation with a real case and give an outlook on performance.

## 1 INTRODUCTION

With the increasing use of graph databases for business- and privacy-critical applications, not only the continuous growth of data and its complexity must be considered but also advanced, flexible, and fine-grained authorization and access control. Access control protects assets and private information against unauthorized access by potentially malicious parties (Benantar, 2005). Authorization is the process and result of specifying access rights in terms of who (subject) can perform what (action) on which resource (Jøsang, 2017). An authorization policy defines access rights in one or more sets of rules using some policy language. Existing policy languages do not yet consider graph-specific characteristics.

A graph is a set of vertices which can be related to each other in pairs by edges. In graph databases, both vertices and edges are stored and accessed as entities. Thus, vertices can be also considered in the context of their relationships to other vertices, even over longer paths, i.e. sequences of alternating vertices and edges.

Consider a knowledge graph with *data objects* and *tasks* as vertex entities. We need to describe a path in the authorization policy with constraints on the attributes of subject and resource *data object* vertices. Moreover, a *task* vertex has to exist somewhere along this path with a connecting edge of certain values for its *typeCode* attribute.

We worked on attribute-level path constraints in previous research iterations (see Section 3.3) resulting in initial versions of our model called *XACML for Graphs (XACML4G)*. The policy language and access control model support constraints on paths, but each element of the path must be described in detail. Furthermore, only graphs in ArangoDB are supported.

Our current research deals with the open issues from these earlier iterations as well as further input from related work. We contribute the following results to the XACML4G model and prototype:

- Flexible path specification in XACML4G authorization policies, without defining every vertex and edge in the path pattern.

- Edges are also considered as resources.

---

[a]  https://orcid.org/0000-0001-8972-625
[b]  https://orcid.org/0000-0001-5094-2248
[c]  https://orcid.org/0000-0003-0310-1942
[d]  https://orcid.org/0000-0002-9858-837X

- A database-independent enforcement model using a source-subset graph.
- Support path-related attributes in the XACML4G policy and request by extending an established XACML implementation.
- A proof of concept prototype of the extended XACML4G language and architecture, implementing flexible authorization policy specification and datastore-independent enforcement.
- A demo case to show the extended XACML4G policy definition and enforcement as well as extended XACML request handling.

The rest of the paper is structured as follows. In Section 2, we provide an overview of our research method and research questions. Section 3 gives an overview of relevant access control models, technologies, and results of our previous research iterations. The policy language and enforcement details of our current XACML extension are explained in Section 4. Section 5 gives details about our prototypical implementation and a demo case. In Section 6, we investigate the performance of the implemented prototype compared with XACML and our previous work. The paper concludes with a summary and an outlook on future work in Section 7.

Concerning our DSR knowledge base, we discuss in Section 3 the results of our previous design iterations as well as approaches and technologies. The focus of our current research iteration is on the following research questions (RQ):

**RQ1** What are the main challenges for a more flexible definition and datastore-independent enforcement of XACML4G path constraints?

**RQ2** What are suitable concepts to design a more flexible definition and datastore-independent enforcement of XACML4G path constraints?

**RQ3** Can a prototypical implementation of the proposed concept be provided and applied to real access control cases?

We use different methods for the evaluation of our design. In the first two research questions, we focused on analytical proofs and feedback circles with our project partners. To address *RQ3*, we implemented a proof of concept prototype showing its feasibility on a generated dataset from the domain of patent and trademark prosecution in a patent law firm (see Section 5). In this paper, we overcome the limitations of the previous iterations concerning flexible definition of path constraints and enforcement in a generally applicable way.

## 2 Research Method

We follow a design science research (DSR) method (Hevner et al., 2004; vom Brocke et al., 2020) and thus, focus on problem solving to enhance human knowledge by designing artifacts. The requirements for authorization and access control in the context of graph-structured data originate from a real-world problem in the domain of IT-supported knowledge work in a patent law firm (Hübscher et al., 2021).

Concepts, prototypes and knowledge are designed, developed and evaluated in an iterative research process within two complementary projects with partners from business and research. While the first two iterations were strongly driven by the needs of our business partner, we generalize the concept in this iteration to be applicable in a broader context, i.e., flexible, adaptable and datastore-independent.

The initial scope of our research was in the area of highly flexible, knowledge-intensive business processes for patent and trademark prosecution. We designed a dynamic fine-grained authorization and access control solution for a knowledge graph integrating processes and data, which is currently implemented in the multi-model database ArangoDB.

## 3 Related Work

We provide an overview of the knowledge base relevant to our current research iteration. Besides the related access control models to protect graph-structured data, we introduce the models and technological background as well as the main results of our previous research iterations that significantly drive our work.

### 3.1 Access Control Models

Currently, graph and multi-model database systems focus on *role-based access control (RBAC)*, e.g., *Neo4j* (Borojevic, 2017), *Microsoft Azure CosmosDB* (Brown et al., 2021), and *ArangoDB* (ArangoDB, 2022), where authorizations are defined for each entity individually. However, RBAC is not sufficient as it neither considers protecting data via restrictions on paths in the graph nor takes entity properties into account (see Section 4.1).

To protect graph-structured data, the path from the subject to the resource needs to be constrained not only by properties such as depth, type or trust level as shown in several relation-based access control (ReBAC) models, but also by content. ReBAC is an access control paradigm based on evaluating relationships between subjects requesting access and resources. During the last decade different ReBAC models emerged (Giunchiglia et al., 2008; Fong, 2011; Cheng et al., 2016). However, there is no common definition of ReBAC which resulted in a number of domain-specific models with rather ad-hoc enforcement models and implementations (Clark et al., 2022). Clark et al. consider ReBAC policies as graph queries over graph databases. They compared representative ReBAC models and derived ReBAC policy language features, which are formalized in their ReBAC query language *ReLOG*.

Most ReLOG features are already supported in our previous XACML4G versions, such as querying basic graph patterns, mutual exclusion constraints, arbitrary path semantics, path negation, parameterized queries, or the any predicate. Path patterns do not need to describe the overall path from subject to resource in ReLOG. This is an open issue in previous XACML4G versions, but we consider it in the current research iteration. ReLOG is based on the *regular property-graph logic* introducing custom functions to encode edge types and overcome the limited expressiveness of the policy language concerning path constraints. When the policy is changed, new functions need to be implemented leading to a a huge amount of functions, which are hardly manageable.

Fine-grained access control and an implementation are still open issues with ReLOG. As we follow a design science research approach, implementing our approach and testing it on cases from practice has always been an important part of our research process. We identified an open issue for ReLOG and XACML4G: only vertices are considered as resources. However, as edges are also core elements in graphs, they need to be considered as protectable resources too.

Neither RBAC nor comprehensive ReBAC models (e.g., ReLOG) support fine-grained access control applying constraints on the entities to be protected. Attribute-based access control (ABAC) (Hu et al., 2017) overcomes this limitation. Constraints can be defined at the attribute level for subjects, resources, actions to be performed, and environment conditions. Unlike ReBAC, the ABAC model lacks the natural specification of relationships between the subject and resource. The ABAC and ReBAC models are compared in (Ahmed et al., 2017). They analyzed a family from each model showing that ABAC models are generally more expressive than ReBAC models (Rizvi and Fong, 2020). In the current research iteration, we proceed with the development of our approach integrating ReBAC features with ABAC.

## 3.2 Models and Technologies

Braun et al. (Braun et al., 2008) regard XML-based models such as XACML (see Section 3.2.1) as the closest to graph-related requirements. Thus, we rely on XACML as it is considered the defacto standard for managing and enforcing fine-grained privileges such as the *PRIMA* system in Markus et al. (Lorch et al., 2003), Han (Tao, 2005), Jin et al. (Wu et al., 2006), and many other works. As our current design relies on a graph database to enforce XACML4G policies, we further discuss the graph query language Cypher and the graph database Neo4j.

### 3.2.1 XACML

XACML is the abbreviation of *eXtensible Access Control Markup Language*. It is an *OASIS*[1] approved standard for access control established in 2001. The policy language model of XACML is XML-based having the three main components: rule, policy, and policy set. Firstly, *rule* is the basic element having an effect (i.e., permit or deny) as well as an optional target and condition. A *target* is a combination of zero or more subjects, resources, actions, and environment attributes. A *policy* is comprised of zero or more rules, a rule combining algorithm, and a target. Last but not least, a *policy set* is a composite element consisting of a target, a policy combining algorithm besides zero or more policy sets and policies. A rule, policy, or policy set is applicable when its target attributes match those in the request.

XACML is not only a policy language, but also a processing model (i.e., architecture, workflow, and methodology) for evaluating access requests. The data flow between the XACML conceptual units is visualized in Figure 1. Additionally, XACML provides extension points for defining custom combining algorithms, attribute providers, policy providers, data types, and functions.

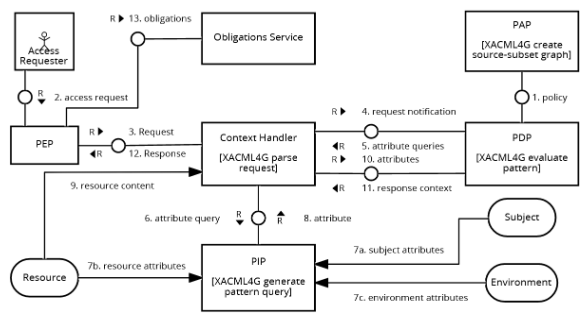

Figure 1: XACML reference architecture and extension

---

[1] http://www.oasis-open.org

The *policy administration point (PAP)* manages policies, which will be used in evaluating access requests, with respect to authoring and deployment (1). The *policy enforcement point (PEP)* receives the access request from the user (2), maps it to the XACML request native format and sends it to the context handler (3). Furthermore, the PEP fulfills obligations, i.e., operations carried out during the policy enforcement phase (13). The context handler converts access requests from the native format to the XACML canonical form (4) and vice versa for the response (12). It also acts as an intermediate entity between the *policy decision point (PDP)* and the *policy information point (PIP)*. The PDP requests subject, resource, action, environment, and other custom attributes from the context handler (5). The context handler requests the attributes from the PIP (6), retrieves them from the respective entities (7) and returns them to the context handler (8). The context handler further optionally includes the resource in the context (9). Finally, the results are sent to the PDP (10) for evaluating the policies and making authorization decisions (11).

#### 3.2.2 Cypher Query Language

Different query languages are proposed for property graphs (i.e., directed, labeled, attributed multi-graphs (Fletcher et al., 2018)) such as PGQL (van Rest et al., 2016), Gremlin[2], Blueprints[3], G-CORE (Angles et al., 2018), and Cypher[4,5]. In addition, a standardized declarative query language for property graphs, comparable to SQL for relational data, is being developed in the ongoing ISO project *Graph Query Language (GQL)*[6].

Cypher is a declarative query language for property graphs and has a syntax inspired by SQL. It allows to easily express graph patterns as well as path queries. Cypher was originally created by Neo4j and contributed to the open-source project openCypher in 2015. Therefore, it is not only used with Neo4j, but also available with other graph database systems[7] such as Amazon Neptune or SAP HANA Graph. Due to its expressiveness, flexibility, and significant contribution to the design and development of the future ISO standard GQL, we decided for Cypher as the most suitable declarative graph query language for processing and evaluating patterns in XACML4G authorization policies.

---

[2] https://tinkerpop.apache.org/gremlin.html
[3] https://github.com/tinkerpop/blueprints/wiki
[4] https://neo4j.com/developer/cypher/
[5] https://opencypher.org/
[6] https://www.gqlstandards.org/
[7] https://opencypher.org/projects/

#### 3.2.3 Neo4j

Neo4j (Neo4j, 2022) is a native graph database implementing the property graph model in Java. The data is structured in terms of nodes and relationships. It supports schema-free and schema-optional use. It is currently available in the open source Community Edition (GPL v3) and the commercial Enterprise Edition. In over 20 years, they established their position as world market leader in graph databases[8] with significant evolution and an active community.

The Neo4j database is directly accessed and queried using the declarative query language *Cypher*. Furthermore, it provides means to load data from different sources, which is needed for implementing our demo cases. Thus, we decided to use Neo4j to process and evaluate XACML4G policies.

### 3.3 Previous Results

In the following, we summarize the results of our previous research iterations to develop an attribute-based access control model for graph-structured data considering path constraints.

In (Mohamed et al., 2021a), we presented a preliminary approach to define and enforce graph-specific authorizations. We proposed a model for expressing fine-grained constraints on vertex and edge properties along the path between a subject and resource. We introduced a JSON-formatted authorization policy based on the XACML policy structure and provided a proprietary pattern enforcement. The concept and prototype are restricted to the multi-model database *ArangoDB*.

Our subsequent work (Mohamed et al., 2021b) proves the concept in the context of XACML. We provide a formal grammar for the extended XACML policy and request. Furthermore, the reference architecture of XACML is extended to enforce the newly introduced element (i.e., *pattern*) using the extensibility points of the standard functional components. This XACML extension has the expressiveness of standard XACML in addition to considering path constraints for graph-structured data. However, the complete path between a given subject and resource needs to be defined. Moreover, the policy is processed in advance to generate a query for each rule having a pattern and a XACML condition associated to the pattern identifier to evaluate the pattern in the request evaluation phase. This is because the pattern is an extension and cannot be evaluated directly by the XACML model. Requests are also processed to extract the subject and resource from the path attributes to be used in the policy matching by XACML. The other attributes are used for the pattern evaluation. This model is based on ArangoDB and its query language *AQL* and thus, needs further implementation to be used with other graph databases.

---

[8] https://db-engines.com/en/ranking/graph+dbms

## 4 Approach

In this section, we illustrate the extended XACML4G policy and request language including a formal definition of the introduced elements. We explain the proposed extensions of the XACML architecture, whether proprietary or using extensibility points. Moreover, we discuss the policy processing and decision procedures on the conceptual level. We provide the source code of our prototype, an XML schema definition (XSD) of the new elements, and a demo case in our documentation[9].

### 4.1 Challenges

Based on the results of our previous research iterations and related work, we identify the following challenges:

- Flexible number of hops between two path vertices without specifying every path element.
- Pattern-related conditions (e.g., joining conditions between path elements).
- A datastore-independent enforcement model of XACML4G policies for property-graph compatible datastores.

We now present the latest XACML4G version addressing the limitations of our previous results.

### 4.2 Policy Language Extension

In the following, we describe the policy, rule, and request structure of XACML4G.

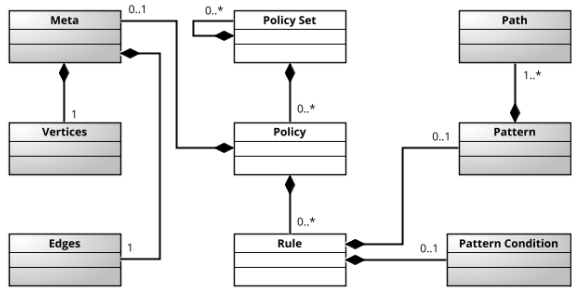

Figure 2: XACML4G policy language model

---
[9]XACML extension for graphs documentation

### 4.2.1 XACML4G Policy

The XACML policy is extended with the new element *Meta* as depicted in Figure 2. The *meta* element defines the vertices and edges relevant to evaluate the policy. We manage this subset of the overall (source) data in an independent graph, which we call, according to its purpose, *source-subset graph*.

The meta element is composed of *Vertices* and *Edges*, where each of them lists at least one label and type value for nodes and relationships of the source graph respectively.

Listing 1 represents the XSD of the meta element.

Listing 1: XSD for the meta element
```
<xs:element name="Meta">
  <xs:complexType><xs:sequence>
    <xs:element name="Vertices">
      <xs:complexType><xs:sequence>
        <xs:element name="VertexEntity" minOccurs="1"
          maxOccurs="unbounded" type="xs:string"/>
      </xs:sequence></xs:complexType>
    </xs:element>
    <xs:element name="Edges">
      <xs:complexType><xs:sequence>
        <xs:element name="EdgeEntity" minOccurs="1"
          maxOccurs="unbounded" type="xs:string"/>
      </xs:sequence></xs:complexType>
    </xs:element>
  </xs:sequence></xs:complexType>
</xs:element>
```

### 4.2.2 XACML4G Rule

The XML schema of the XACML rule is extended by adding two elements, *Pattern* and *PatternCondition* (see Figure 2). The pattern element is already introduced in previous iterations as a path with recursive structure consisting of a vertex, an edge, and either another vertex (i.e., the base case) or an entire path. Although the pattern definition does not change, the elements within the path are enhanced to support flexible patterns. Thus, there is no need to specify every vertex and edge along the path anymore.

The path vertex (cp. Listing 2) and edge elements express the pattern constraints by specifying the attributes and their values as a sequence of *AnyOf*, like in the standard XACML target. The optional attribute *Label* is added to specify the entity/collection to which the vertex belongs. We consider the label as part of the vertex definition, not a property to be matched against a value. The vertex has an attribute *Category* indicating whether it is a subject, a resource, or belongs to the path. The subject and resource categories are predefined by XACML while the path category is defined in the extended version to handle the path vertices differently. The *VertexId* attribute represents the identifier of the vertex element. The identifiers are mainly used to join vertices or edges in the pattern condition element.

Listing 2: XSD for the path vertex

```xml
<xs:element name="Vertex">
  <xs:complexType>
    <xs:sequence minOccurs="0" maxOccurs="unbounded">
      <xs:element ref="xacml:AnyOf"/>
    </xs:sequence>
    <xs:attribute name="VertexId" type="xs:string"
     use="optional"/>
    <xs:attribute name="Label" type="xs:string"
     use="optional"/>
    <xs:attribute name="Category" use="optional">
      <xs:simpleType>
        <xs:restriction base="xs:string">
          <xs:enumeration value="urn:oasis:names:tc:
           xacml:1.0:subject-category:access-subject"/>
          <xs:enumeration value="xacml4g:1.0:path-
            category:vertex"/>
          <xs:enumeration value="urn:oasis:names:tc:
           xacml:3.0:attribute-category:resource"/>
        </xs:restriction>
      </xs:simpleType>
    </xs:attribute>
  </xs:complexType>
</xs:element>
```

The edge XSD is shown in Listing 3. The constraints within an edge element are structured like in the vertex. The edge can belong to the path or resource category only. The edge element is identified by the *EdgeId* attribute. The direction of the edge (e.g., inbound, outbound, or any) can be indicated by the attribute *Direction*. The *Length*, *MinLength*, and *MaxLength* attributes add flexibility to the pattern by specifying a range for some part of the path. The *Type* attribute is similar to the vertex label, thus indicating the edge type.

Listing 3: XSD for the path edge

```xml
<xs:element name="Edge">
  <xs:complexType>
    <xs:sequence minOccurs="0" maxOccurs="unbounded">
      <xs:element ref="xacml:AnyOf"/>
    </xs:sequence>
    <xs:attribute name="EdgeId" type="xs:string"
     use="optional"/>
    <xs:attribute name="Type" type="xs:string"
     use="optional"/>
    <xs:attribute name="MinLength" type="xs:integer"
     use="optional"/>
    <xs:attribute name="MaxLength" type="xs:integer"
     use="optional"/>
    <xs:attribute name="Length" type="xs:integer"
     use="optional"/>
    <xs:attribute name="Category" use="optional">
      <xs:simpleType>
        <xs:restriction base="xs:string">
          <xs:enumeration value="xacml4g:1.0:path-
            category:edge"/>
          <xs:enumeration value="urn:oasis:names:tc:
           xacml:3.0:attribute-category:resource"/>
        </xs:restriction>
      </xs:simpleType>
    </xs:attribute>
    <xs:attribute name="Direction" use="optional">
      <xs:simpleType>
        <xs:restriction base="xs:string">
          <xs:enumeration value="from"/>
          <xs:enumeration value="to"/>
          <xs:enumeration value="any"/>
        </xs:restriction>
      </xs:simpleType>
    </xs:attribute>
  </xs:complexType>
</xs:element>
```

*Pattern condition* is another new element. It describes constraints and joining conditions that are related to the path elements of the pattern within the same rule. It consists of a sequence of conditions in the XACML format as defined in Listing 4. The elements in the sequence within the pattern condition have the same structure as the XACML *Apply* element, but its *AttributeDesignator* element is extended to include a *VertexId* or an *EdgeId* attribute representing the variable defined for a vertex or an edge in the rule pattern. Accordingly, the category should be either *xacml4g:1.0:path-category:vertex* or *xacml4g:1.0:path-category:edge*.

Listing 4: XSD for the pattern condition

```xml
<xs:element name="PatternCondition">
  <xs:complexType><xs:sequence>
    <xs:element ref="xacml:Apply"/>
  </xs:sequence></xs:complexType>
</xs:element>
```

Furthermore, the function attribute (i.e., *FunctionId*) of the *Apply* element should be one of the predefined XACML4G functions. For now, we support conjunction, disjunction and comparison operators besides some string functions as illustrated in Table 1. The function URI is then translated to the corresponding operation in Cypher during the dynamic query generation in the request evaluation phase.

Table 1: XACML4G functions

| Function | URI |
| --- | --- |
| AND | xacml4g:1.0:function:and |
| OR | xacml4g:1.0:function:or |
| Equal | xacml4g:1.0:function:equal |
| Greater than ($>$) | xacml4gx:1.0:function:greater-than |
| Greater than or equal ($\geq$) | xacml4g:1.0:function:greater-than-or-equal |
| Less than ($<$) | xacml4g:1.0:function:less-than |
| Less than or equal ($\leq$) | xacml4g:1.0:function:less-than-or-equal |
| Not equal (!=) | xacml4g:1.0:function:not-equal |
| String equal case sensitive | xacml4g:1.0:function:string-equal-ignore-case |
| String contains | xacml4g:1.0:function:string-contains |
| String starts with | xacml4g:1.0:function:string-starts-with |

### 4.2.3 XACML4G Request

The standard XACML access request consists of a sequence of attributes related to the subject, resource, and action. Each request attribute has an attribute id, a data type and a value. In (Mohamed et al., 2021b), we extended the XACML request to differentiate between the attributes' types, i.e., action and path attributes. Both types of attributes have the same structure as the ordinary XACML attributes. The path attributes are needed in matching requests against existing policies, especially subject and resource attributes, as well as during the pattern evaluation in the PIP. Therefore, we defined a *path* category to which all vertices other than subject and resource vertices belong.

In this work, we are extending the request with respect to language and its processing. The action and path elements are defined as a sequence of the standard XACML *Attributes* consisting of an attribute element having an identifier (i.e., *AttributeId*) as a property and an element for the value (i.e., *AttributeValue*).

The definition of the *Attributes* element is extended to optionally include an attribute called *Type* for specifying whether it represents a vertex or an edge (see Listing 5 below).

Listing 5: XSD of the XACML *Attributes* element

```xml
<xs:element name="Attributes" type="AttributesType"/>
  <xs:complexType name="AttributesType">
    <xs:sequence>
      <xs:element ref="Content" minOccurs="0"/>
      <xs:element ref="Attribute" minOccurs="0"
        maxOccurs="unbounded"/>
    </xs:sequence>
    <xs:attribute name="Category" type="xs:anyURI"
      use="required"/>
    <xs:attribute name="Type" use="optional">
      <xs:simpleType>
        <xs:restriction base="xs:anyURI">
          <xs:enumeration value="xacml4g:1.0:path-
              category:vertex"/>
          <xs:enumeration value="xacml4g:1.0:path-
              category:edge"/>
        </xs:restriction>
      </xs:simpleType>
    </xs:attribute>
    <xs:attribute ref="xml:id" use="optional"/>
</xs:complexType>
```

For the path attributes, we need to define not only the category, entity, and value for a vertex (or a resource edge), but also to which property this value belongs. Therefore, we provide the property name and value in the *AttributeValue* tag separated by a colon. We also use this format to specify the identifier name and value of a vertex or a resource edge in the request since the naming convention of the identifier property could vary from a data source to another.

Several components of the XACML architecture, i.e., context handler, PAP, PIP and PDP, are extended as illustrated in Figure 1 to handle the language-specific extensions according to the proposed enforcement concept.

### 4.3 Architecture Extension

To apply the XACML4G language extensions, the policy enforcement model is extended to deal with paths in the XACML4G requests and match them against the patterns within rules of the XACML4G policy evaluating the pattern conditions as well. We propose a datastore-independent enforcement concept to evaluate the policy from various data sources without changing the core model. Therefore, we introduce an optional property graph, which we call *source-subset graph*. It contains all data from the source datastore needed to evaluate the authorization policies.

Firstly, the policy administration point (PAP) is extended to parse the policy files and extract the meta element, which is used to create the source-subset graph. The values of the *VertexEntity* and *EdgeEntity* elements (see Listing 1) represent the node labels and relationship types in graph databases such as Neo4j. The source-subset graph can be created from multiple data sources including flat files or database systems. The only requirement is that the data model can be mapped to the property graph model. Creating the source-subset graph is independent of the request evaluation and can be updated in the background. If our model is configured to directly interact with the source datastore and not to use the optional source-subset graph, then the *XACML4G create source-subset graph* phase (refer to Figure 1) is irrelevant.

The context handler receiving the access requests is extended to parse the path and action attributes, which are used in policy matching by XACML as well as generating a Cypher pattern corresponding to the request path attributes by XACML4G. This is done by extending an established open source XACML implementation (see Section 5.1), to extract the structured attributes and use the path attributes in the policy evaluation phase.

The policy information point (PIP) is extended to automatically handle the policy attributes related to the vertex label and edge type as well as the custom attributes representing the pattern identifier of the rule being evaluated. A Cypher pattern and a *where* statement are dynamically generated from the pattern and pattern condition elements to evaluate the pattern-related attributes (refer to *XACML4G generate pattern query* extension in the PIP component in Figure 1). Then, the result is added to the Cypher query template in Listing 6. This query searches for the intersection of the rule pattern from the matched policy (p1) and request pattern (p2). The request pattern is generated from the path attributes representing the input path in the request.

Listing 6: Cypher query template for matching the rule pattern and its conditions with the request path

```
MATCH p1 = <rule pattern of matched policy>
MATCH p2 = <pattern of request path attributes>
WHERE <pattern condition(s) of matched policy rule>
  AND ALL (x IN nodes(p2) WHERE x IN nodes(p1))
  AND ALL (x IN relationships(p2)
      WHERE x IN relationships(p1))
RETURN p1 IS NOT NULL AS result
```

When the target of the policy is matched with the subject, resource, action, and environment attributes in the request, the XACML policy decision point (PDP) proceeds with evaluating the policy rules to determine the access decision. Before evaluating the conditions of the matched policy rules, we append an additional XACML condition for the rules having a pattern. This condition is specific to evaluating the XACML4G pattern and its conditions (see *XACML4G evaluate pattern* in the PDP in Figure 1). The condition evaluation is successful if the query returns a result (i.e., true). The query is generated in the PIP according to the pattern and its conditions within the rule. It is executed in a Neo4j database, which contains the source-subset graph (either the source datastore or the additional optional source-subset graph). If the condition fails to evaluate due to the pattern query, an indeterminate decision is returned.

We now demonstrate details about our prototypical implementation and a case from practice.

## 5 DEMONSTRATION CASE

For demonstrating the feasibility of the proposed approach, we implement the proposed concept and apply it to real access control cases. In this paper, we present an access control scenario from one of our research projects. We start with an overview of the data model and the selected scenario. The steps, intermediate results, and the final decision are presented from the policy formulation and processing till the final decision of evaluating an access request including a conflict case.

### 5.1 Implementation

Our prototype is implemented in Java and we use Neo4j along with its query language Cypher to enforce the authorizations. Authorization-relevant data is either stored in a Neo4j source datastore or the optional Neo4j source-subset graph. XACML4G is not restricted to a certain database. According to the source datastore, a respective importer class connects to the source datastore, retrieves the required data from the defined vertex and edge entities of the meta element in the policy, and creates the source-subset graph. Currently, we implemented two Java classes to import graph data for Neo4j and ArangoDB. It is also possible to import data from other datastores or even flat files (e.g., CSV and JSON) by simply implementing a class for that custom data source. The created source-subset graph is queried in the request evaluation phase independently of the source datastore. XACML4G only operates on the relevant subgraph according to the established policy.

The prototype is based on the open-source XACML implementation *Balana*[10]. The XACML4G policy and request are validated using the XML schema in Section 4.2. The policy is expressed in the standard XACML syntax and the request evaluation against the stored policies is handled by the Balana framework. We extended the source code of Balana as follows:

[10]https://github.com/wso2/balana

- Parse the structured attributes (i.e., action and path) in XACML4G requests.
- Evaluate the XACML4G pattern (if exists) by adding a XACML condition dynamically during evaluating the rule(s) of the matched policy (or policies) against the access request.

Our extensions address the open issues stated in Section 4.1 without affecting the overall XACML-specific procedures and results. The implementation is demonstrated by a real case in the following section.

### 5.2 TEAM Model Case

Since the first iterations to design our approach XACML4G we were part of the KnoP-2D project. Then, we have continued to use the TEAM model (Hübscher et al., 2021) case in patent and trademark prosecution to demonstrate XACML4G in a real case on the conceptual and implementation level. The TEAM model follows a graph-based, meta-modeling approach. It consists of three layers: the meta model, the domain model, and the instance model (Hübscher et al., 2020).

For simplicity, we only focus on the subgraph relevant to our access control scenario in the instance model, i.e., the graph containing all relevant instances from the real world case. A vertex collection called *dataObjects* has entities that are related to each other via edges named *dataObjectRelations* or connected to *tasks* through *accessRelations* and *taskDataRelations*. The *accessRelations* express user-to-task relationships whereas *taskDataRelations* link *tasks* to either *dataObjects* or *dataObjectRelations*.

Due to privacy reasons, we used a generated test dataset. The graph model contains 11,982 dataObjects and 2,559 tasks as vertices in addition to 3,165 accessRelations, 13,271 dataObjectRelations, and 53,246 taskDataRelations as edges. It is originally imported into the multi-model database *ArangoDB* where each entity class is represented as a vertex or edge collection. An edge collection entry requires the identifier of the source and destination collection entries. We define a scenario specific to the instance layer of the TEAM graph model in Example 1:

**Example 1.** *As a user, I can access a data object if I am allocated to or work/have worked on a task that has a connecting path to the resource.*

The rule components of the authorization policy for this example are listed below. The + symbol indicates a path with variable length (1 or more) from the task to the resource.

- *Subject*: user
- *Resource*: data object
- *Action*: access
- *Pattern*: user → task →$^+$ data object
- *Effect*: permit

After establishing the access control scenario, the authorization policy is defined according to the XACML4G policy format in Section 4.2.1. The defined policy is illustrated in Listing 7. The information given in the meta element defines the vertex and edge collections of the TEAM graph model, which are relevant for creating the source-subset graph.

The subject user is represented as a data object having *typeCode* and *pmUser* as an attribute key and value respectively whereas the resource could be any data object. For expressing the pattern, the subject and resource defined in the target are the start and end vertices respectively. We then specify the task as an intermediate vertex between the user connected via an access relation edge and a maximum of two hops away from the resource. The pattern in this example shows how to express attribute-based constraints on path elements (i.e., vertices and edges) with the option to flexibly specify the minimum and/or maximum range of a part of the path. For the pattern condition element, we define a variable for an edge of type *accessRelations* and specify constraints for the *typeKind* property of this variable.

Listing 7: XACML4G policy for the TEAM case

```
<xacml:Policy PolicyId="pmUserToDataObject"
 RuleCombiningAlgId="urn:oasis:names:tc:xacml:1.0:rule-
      combining-algorithm:first-applicable">
 <xacml4g:Meta>
  <xacml4g:Vertices>
   <xacml4g:VertexEntity>dataObjects</xacml4g:VertexEntity>
   <xacml4g:VertexEntity>tasks</xacml4g:VertexEntity>
  </xacml4g:Vertices>
  <xacml4g:Edges>
   <xacml4g:EdgeEntity>dataObjectRelations
   </xacml4g:EdgeEntity>
   <xacml4g:EdgeEntity>accessRelations</xacml4g:EdgeEntity>
   <xacml4g:EdgeEntity>taskDataRelations</
        xacml4g:EdgeEntity>
  </xacml4g:Edges>
 </xacml4g:Meta>
 <xacml:Rule Effect="Permit" RuleId="user_access_dataObj">
  <xacml4g:Pattern PatternId="userToDataObjectAccess">
   <xacml4g:Path>
    <xacml4g:Vertex Category="urn:oasis:names:tc:xacml:1.0
         :subject-category:access-subject" VertexId="s">
     <xacml:AnyOf>
      <xacml:AllOf>
       <xacml:Match MatchId="urn:oasis:names:tc:xacml:1.0
            :function:string-equal">
        <xacml:AttributeValue>pmUser</xacml:AttributeValue>
        <xacml:AttributeDesignator AttributeId="typeCode"
         Category="xacml4g:1.0:path-category:vertex"/>
       </xacml:Match>
      </xacml:AllOf>
     </xacml:AnyOf>
    </xacml4g:Vertex>
    <xacml4g:Edge Type="accessRelations" Direction="from"
         EdgeId="e" Category="xacml4g:1.0:path-category:edge"/>
    <xacml4g:Path>
     <xacml4g:Vertex Label="tasks"
      Category="xacml4g:1.0:path-category:vertex"/>
     <xacml4g:Edge MaxLength="2"
      Category="xacml4g:1.0:path-category:edge"/>
     <xacml4g:Vertex Category="urn:oasis:names:tc:xacml:3.0
          :attribute-category:resource"/>
    </xacml4g:Path>
   </xacml4g:Path>
  </xacml4g:Pattern>
  <xacml4g:PatternCondition>
   <xacml:Apply FunctionId="xacml4g:1.0:function:or">
    <xacml:Apply FunctionId="xacml4g:1.0:function:equal">
     <xacml:AttributeDesignator AttributeId="typeKind"
      Category="xacml4g:1.0:path-category:edge" EdgeId="e"/
           >
     <xacml:AttributeValue>worksOn</xacml:AttributeValue>
    </xacml:Apply>
    <xacml:Apply FunctionId="xacml4g:1.0:function:equal">
     <xacml:AttributeDesignator AttributeId="typeKind"
      Category="xacml4g:1.0:path-category:edge" EdgeId="e"/
           >
     <xacml:AttributeValue>allocates</xacml:AttributeValue>
    </xacml:Apply>
   </xacml:Apply>
  </xacml4g:PatternCondition>
 </xacml:Rule>
</xacml:Policy>
```

Upon defining the authorization policy, the source-subset graph can be created from the information in the meta element. In the following, we illustrate the steps of evaluating a XACML4G request, as depicted in Listing 8, against the policy defined for the selected scenario. Specifying the attributes related to the subject, resource, path vertices with respect to id (i.e., subject, resource and path vertex) and value are not sufficient. Hence, we describe the *AttributeValue* in the path attributes in terms of name and value separated by a colon.

Listing 8: XACML4G request for the TEAM case

```
<Request ReturnPolicyIdList="true">
 <xacml4g:ActionAttributes>
  <Attributes Category="urn:oasis:names:tc:xacml:3.0
       :attribute-category:action">
   <Attribute AttributeId="urn:oasis:names:tc:xacml:1.0
        :action:action-id">
    <AttributeValue>access-do</AttributeValue>
   </Attribute>
  </Attributes>
 </xacml4g:ActionAttributes>
 <xacml4g:PathAttributes>
  <Attributes Category="urn:oasis:names:tc:xacml:1.0
       :subject-category:access-subject">
   <Attribute AttributeId="urn:oasis:names:tc:xacml:1.0
        :subject:subject-id">
    <AttributeValue>_key:1196741133</AttributeValue>
   </Attribute>
  </Attributes>
  <Attributes Category="xacml4g:1.0:path-category:vertex">
   <Attribute AttributeId="xacml4g:1.0:path:vertex-id">
    <AttributeValue>_key:1196741778</AttributeValue>
   </Attribute>
  </Attributes>
  <Attributes Category="urn:oasis:names:tc:xacml:3.0
       :attribute-category:resource">
   <Attribute AttributeId="urn:oasis:names:tc:xacml:1.0
        :resource:resource-id">
    <AttributeValue>_key:1196742142</AttributeValue>
   </Attribute>
  </Attributes>
 </xacml4g:PathAttributes>
</Request>
```

If the request is matched with a policy having a pattern in its rule(s), a custom XACML condition is added to the conditions list of this matched policy rule. This condition checks whether the query associated to the pattern identifier returns true or not.

Then, the extended PIP generates the pattern query from (1) the pattern and pattern condition in the XACML4G rule, and (2) the path attributes of the XACML4G request. The input path in the XACML4G request is just a sequence of vertices except for the resource, which can also be an edge.

Thus, the edges of the request path pattern have no direction. The request path is only matched by its attributes such as type, category, etc. The direction defined in the rule pattern is applied to the request during the evaluation of the pattern query.

Listing 6 shows the final query according to the Cypher query template.

```
MATCH p1 = (s:dataObjects{typeCode:"pmUser"})-[e1:
   accessRelations]->(:tasks)-[*..2]-(:dataObjects)
MATCH p2 = ({_key:"1196741133"})-[]-
   ({_key:"1196741778"})-[]-({_key:"1196742142"})
WHERE e1.typeKind="worksOn" OR e1.typeKind=
   "allocates" AND ALL (x IN nodes(p2) WHERE x IN
   nodes(p1)) AND ALL (x IN relationships(p2) WHERE
   x IN relationships(p1))
RETURN p1 IS NOT NULL AS result
```

The Cypher query for the intersection of rule and request path patterns is successfully evaluated, as we check for an existing path in the provided dataset which satisfies the pattern constraints. The final decision will be *permit* according to the effect of the successfully evaluated policy rule. The response in Figure 3 includes the identifier of the matched policy. The status code *ok* indicates a determined decision without errors.

```
<Response xmlns="urn:oasis:names:tc:xacml:3.0:core:schema:wd-17">
  <Result>
    <Decision>Permit</Decision>
    <Status>
      <StatusCode Value="urn:oasis:names:tc:xacml:1.0:status:ok"/>
    </Status>
    <PolicyIdentifierList>
      <PolicyIdReference>pmUserToDataObject</PolicyIdReference>
    </PolicyIdentifierList>
  </Result>
</Response>
```

Figure 3: XACML response for the TEAM demo scenario

If we specify another rule within the same policy to explicitly deny the resource data object having an attribute _key equal to *1196742142* (the one specified in the request in Listing 8), there will be a conflict in the request evaluation phase. This is because two rules with opposite decisions are matched with the path in the request. In this case, the condition related to the pattern is successfully evaluated for both rules. The rule combining algorithm resolves the conflict, since it occurs within a single policy, and returns *deny*. The rule order matters because we specify the XACML *first applicable* function as a rule combining algorithm. If the new rule is added after the first one, the final decision will be *permit*.

## 6 PRELIMINARY EVALUATION

Besides the demo case in Section 5, we provide preliminary performance measurements for evaluating access requests with different path lengths. Three prototypes are compared implementing the same scenario, which is based on the TEAM model. The first one uses the standard XACML language and architecture. It implements the scenario on top of ArangoDB, but with limitations concerning expressiveness and scalability, because XACML does not support paths. Thus, only subject, object, and action can be defined within the policy and request. Moreover, we used the XACML extensibility point to manually add the custom attributes specified in the policies and statically write the respective queries to be evaluated in the decision-making phase. In this case, each change in the authorization requirements demand adaptations in the policy as well as the implementation. The second prototype is the initially proposed extension of XACML for graph-structured data (Mohamed et al., 2021b) (see Section 3.3). These two prototypes are already investigated in (Mohamed et al., 2021b), but with a more trivial evaluation design. The third prototype is the XACML-based extension discussed in this paper (see Section 5.1).

The evaluation was performed offline on an Intel(R) Core(TM) i7-6500U CPU @ 2.50 GHz with 24 GB RAM. We investigated the execution time of 100 consecutive requests against a defined policy from the request processing phase till receiving the access decision as a response. We performed the experiment three times for each prototype and calculated the average. The results are plotted in Figure 4.

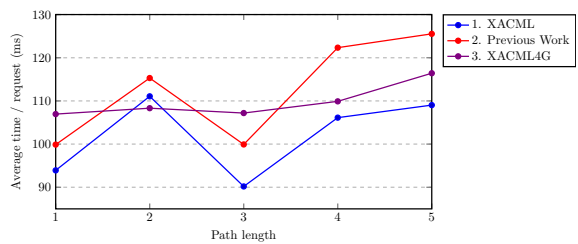

Figure 4: Average request evaluation time for the XACML, previous work, and latest implementation prototypes

The plot for the *XACML* (1) and *previous work* (2) prototypes look like those in (Mohamed et al., 2021b), where we proved that the introduced overhead in the extension is constant regardless of the path length. This overhead almost doubled with prototype (2) compared to (1) as indicated in the difference between the plots for the path length 4 and 5 vs. shorter paths. This is due to the additional pattern conditions to join specific elements along the path. From the implementation perspective, the two major differences between the prototype presented in this paper and the other ones are the underlying graph database and the pattern evaluation within the reference architecture of XACML. Previously, our approach relied on directly connecting to the source graph database and our implementation was based on ArangoDB, the database into which the graph model of the demo case was imported, and its declarative query language *AQL*. In the latest implementation, our approach is independent of the source datastore. It is currently based on Neo4j and its declarative query language *Cypher* using an embedded database within the XACML4G prototype for evaluating patterns and pattern conditions.

As can be observed in Figure 4, our latest work is only slightly impacted with the increased path length. This is most likely due to the better stability of Neo4j compared with ArangoDB. Thus, we plan to repeat the evaluation for all prototypes with the same underlying database to investigate the impact of our extension excluding other factors.

# 7 CONCLUSIONS

In this paper, we presented our enhanced XACML4G language and architecture to consider specific authorization open issues for graph-structured data: flexible constraints on graph paths, protecting edges as a resource, and datastore-independent enforcement. We defined the main challenges for the current research iteration (*RQ1*) in terms of path features and an enforcement model for any property-graph compatible datastore. The path features include support for a flexible number of hops between two path vertices and pattern-related conditions (e.g., joining conditions between path elements). Therefore, not every path element needs to be specified in the pattern.

The second research question (*RQ2*) addresses the concepts to meet these challenges. We selected a declarative graph query language comparable to SQL in relational databases, which supports the required characteristics such as pattern matching on paths, flexible path lengths, or incomplete path specifications. As the ISO standard for the generally applicable graph query language (GQL) on property graphs is still in progress, we rely on Cypher. We proposed using a property graph holding all access control-relevant data (i.e., only a subset of the source graph) independently of the main datastore.

We answered our last research question (*RQ3*) by providing a proof of concept prototype of the extended XACML language and architecture and an access control case for a real knowledge graph. The proof of concept prototype implements our extensions to XACML: (1) extending the XACML language for the policy definition providing a formal grammar in terms of an XML schema definition, (2) and using the extensibility points in the PIP as well as proprietary extensions of the XACML architecture (i.e., context handler, PAP, and PDP) for the policy enforcement. The prototype is implemented in Java, extending the open source XACML implementation *Balana*. Thus, preprocessing of policies and requests (cp. our previous work) is no longer required. *Neo4j* and *Cypher* are used to support datastore-independent policy enforcement. Even though using different datastores for the operational data, all required access restrictions can be defined and enforced without further implementation.

We compared the performance of our prototypical implementation of the current approach with a statically implemented XACML prototype as well as our previous work. The results showed better performance and stability with respect to evaluating paths with different lengths. In contrast to our previous results, the current approach no longer introduces constant overhead.

The current work has highlighted further challenges. Although we managed to integrate the pattern evaluation within the XACML model, it is still evaluated as a XACML condition. Therefore, pattern-related errors cannot be detected. In the future, property graphs with multiple labels on vertices and edges will be considered to match with real-world graph models. To get more reliable performance measurements, we still need to exclude factors affecting the overall performance such as different graph database systems for enforcing the policies. Finally, sophisticated data importers are demanded to construct the source-subset graph, i.e., data import from multiple sources and data masking, especially for sensitive data.

## ACKNOWLEDGMENTS

The research reported in this paper has been partly supported by the LIT Secure and Correct Systems Lab funded by the State of Upper Austria. The work was also funded within the FFG BRIDGE project KnoP-2D (grant no. 871299).